# Stretching graphene using polymeric micro-muscles

F. Colangelo[1,2], A. Pitanti[1], V. Mišeikis[3,4,5], C. Coletti[3,4], P. Pingue[1], D. Pisignano[1,6], F. Beltram[1], A. Tredicucci[1,2,6] & S. Roddaro[1,6]

[1]NEST, Scuola Normale Superiore and CNR Istituto Nanoscienze, Piazza San Silvestro 12, 56127, Pisa (PI), Italy
[2]Fondazione Bruno Kessler (FBK), Via Sommarive 18, 38123 Povo, Trento, Italy
[3]Center for Nanotechnology Innovation @NEST, Istituto Italiano di Tecnologia, Piazza San Silvestro 12, 56127 Pisa, Italy
[4]Graphene Labs, Istituto Italiano di Tecnologia, Via Morego 30, I-16163 Genova, Italy
[5]Consorzio Nazionale Interuniversitario per le Telecomunicazioni (CNIT), Via Moruzzi 1, I-56124 Pisa, Italy
[6]Dipartimento di Fisica, Università di Pisa and CNR Istituto Nanoscienze, Largo B. Pontecorvo 3, 56127 Pisa (PI), Italy

**The control of strain in two-dimensional materials opens exciting perspectives for the engineering of their electronic properties[1–4]. While this expectation has been validated by artificial-lattice studies, it remains elusive in the case of atomic lattices[5]. Remarkable results were obtained on nanobubbles[6] and nano-wrinkles[7], or using scanning probes[8]; microscale strain devices were implemented exploiting deformable substrates[9–13] or external loads[14–17]. These devices lack, however, the flexibility required to fully control and investigate arbitrary strain profiles. Here, we demonstrate a novel approach making it possible to induce strain in graphene using polymeric micrometric artificial muscles (MAMs) that contract in a controllable and reversible way under an electronic stimulus. Our method exploits the mechanical response of poly-methyl-methacrylate (PMMA) to electron-beam irradiation. Inhomogeneous anisotropic strain and out-of-plane deformation are demonstrated and studied by Raman, scanning-electron and atomic-force microscopy. These can all be easily combined with the present device architecture. The flexibility of the present method opens new opportunities for the investigation of strain and nanomechanics in two-dimensional materials.**

Various micro-actuation technologies were explored in recent years, including in particular inorganic microelectromechanical systems[16]. A fascinating alternative approach consists in using polymeric actuators, which exert a force thanks to electrostatic/electrostriction phenomena or to a modification in their oxidation state or molecular conformation. Relevant examples designed through the years include conjugated polymer microrobots actuated by reduction/oxidation cycles for operation in aqueous environments,[18] optically-actuated blends embedding azopolymers or other photochromic dopants,[19] enzymatically-triggered soft components,[20] and shape-memory architectures realized by multimaterial 4D printing.[21] In our case, we exploit PMMA as the MAMs active material. It is indeed well known that high-dose electron radiation can induce cross-linking,[22] leading to a vertical shrinkage up to 50% of thin PMMA films[19,20]. For instance, this effect was used to engineer a strain as large as 24% in silicon nanowires[19]. Here we use e-beam-induced lateral shrinkage of suspended PMMA, a configuration not yet well documented in the literature. A rater unconventional choice as a polymeric actuator, PMMA offers the crucial benefit of being patternable in custom geometries by standard e-beam lithography: this property is relevant for the successful implementation of the MAMs.

Figure 1 illustrates the present sample architecture. First of all, a suspended $Si_3N_4$ membrane, acting as a mechanical support structure, was micro-patterned with an array of circular holes with a 10 μm diameter and then coated with Ti/Au (10/50 nm) to minimize charging effects. Monolayer single-crystal graphene grown by chemical vapour deposition (CVD) was then deposited on top of the substrate using a dry-transfer technique and a PMMA layer as the polymeric vector. In the samples illustrated here (see further results and geometries in the Supplementary Information), our target was to induce a local uniaxial strain. To this end, 2×6 μm² windows were defined by e-beam lithography on each graphene/PMMA membrane. In this way, a region of polymer-free graphene was obtained in the central region of each membrane (see Fig. 1a). PMMA removal frees part of the graphene membrane from the mechanical constraint brought by the polymeric layer. Stretching of the graphene was then obtained by an additional e-beam patterning step, using high-dose exposure (see Fig. 1b). The PMMA layer mechanically contracts when suitably stimulated by electron radiation, mimicking a muscular tissue responding to an electrical stimulus (Fig. 1c). Similarly to a muscle, PMMA does also relax back to its rest state once the stimulus is removed, as discussed in the following. The electrically-actuated polymer thus acts as a MAM that we exploit to stretch the graphene membrane multiple times. We stress that one of the key benefits of this approach is that the MAMs geometry can be fully customized since it is defined by e-beam lithography. This implies that a wide range of different in-plane and out-of-plane deformation profiles can be implemented. Furthermore, the technique is enriched by the possibility of *in situ* direct imaging by scanning electron microscopy (SEM). Further details about the fabrication protocols are reported in the Methods.

In order to demonstrate the action of the MAM and create a local uniaxial strain profile on graphene, we used the sample geometry shown in Fig. 1b: we exposed two rectangular PMMA regions in the proximity of the polymer-free graphene in order to pull it from two opposite sides. Note that the MAMs extend up to the SiN/Ti/Au substrate that provides two rigid clamping points. An image of the sample after electron-beam irradiation is reported in Fig. 1d: the SiN hole is visible as a dark circular region containing a polymer-free graphene window (bright horizontal rectangle) surrounded by suspended graphene/PMMA that can be actuated by selective exposure; the two MAMs are visible as grey vertical rectangles. The central stretched region shows a slightly brighter contrast linked to the formation of nano-wrinkles. This will be discussed in the following in connection with AFM imaging.

The strain induced by the MAMs can be quantitatively analysed by spatially-resolved micro-Raman spectroscopy (µ-Raman). Several spectra were collected, before and after MAM actuation. Figure 1e shows the 2D peak position along a central cross-section (dashed line in Fig. 1d). A large shift is found in correspondence to the pulled region, providing a first evidence of graphene strain. In Fig. 1f, we correlate the G and 2D positions ($\omega_G$ and $\omega_{2D}$ respectively) in order to rule out doping-induced effects[23]. The scatter plot collects all the positions of the Raman signal after irradiation. The overlay lines are the directions along which strain-induced (cyan lines) and doping-induced (grey) shifts are expected. The green spot centred in $(\omega^0_G, \omega^0_{2D}) = (1586.8\pm1.0, 2680.1\pm0.7)$ cm$^{-1}$ indicates peak positions measured on mechanically-relaxed graphene (see Supplementary Information) and the same $\omega^0_{2D}$ value is shown by the dashed green line in Fig. 1e. Following MAM actuation, the Raman signal clearly traces the slope characteristic of strain-induced shifts.

Marked strain anisotropy was demonstrated by a polarized µ-Raman analysis of the G peak (see Fig. 2). The symmetry of the graphene crystal implies that the G peak is two-fold degenerate[9], but this symmetry breaks down when the crystal is subjected to uniaxial strain: the G peak splits into two sub-peaks labeled $G_+$ and $G_-$, with a splitting that depends on the degree of strain anisotropy. The Raman intensities of $G_+$ and $G_-$ depend on the polarization angle of the input laser ($\varphi_l$), of the scattered light ($\varphi_s$) and on the angle between the strain axis and the zigzag graphene crystallographic orientation ($\Delta\varphi$)[9]

$$I(G_-) \propto \sin^2(\varphi_l + \varphi_s + 3\Delta\varphi),$$
$$I(G_+) \propto \cos^2(\varphi_l + \varphi_s + 3\Delta\varphi). \quad (1)$$

By changing $\varphi_l$ or $\varphi_s$ we can modulate the amplitudes of the two G sub-modes that oscillate in antiphase and with a periodicity of π radians. This yields a precise measurement of the strain anisotropy and of the orientation of the graphene lattice. A comparative study is presented here by considering one spot in between the MAMs (yellow point A in Fig. 2a) and one located 2 µm away (orange point B), where the Raman-peak shift is expected to be negligible. Figures 2b and 2c show the G and 2D non-polarized Raman peaks after MAMs actuation. The orange and yellow colored peaks correspond to the sampling positions A and B cited above, respectively. As suggested by the 2D peak shift (see Fig. 2b), graphene is strained in A and barely deformed in B: the 2D peak resonance in unstrained graphene is observed at 2680.1 cm$^{-1}$ (vertical dashed line). Using the Grüneisen parameters[15], the maximum shift we obtained corresponds to a strain of about 0.8% at A (see Supplementary Information). In Fig. 2c, the G peak displays a more complex behaviour. While the spectrum in B is well fitted with a single Lorentzian, the peak at point A shows a pronounced shoulder and can be fitted by a double Lorentzian, as expected for uniaxially strained graphene.

Polarized µ-Raman makes it possible to analyze the nature of the induced strain. We changed $\varphi_l$ while keeping all the other angles fixed and recorded the relative intensities of the $G_{+/-}$ peaks at A. As expected, this leads to a modulation of the $G_+$ and $G_-$ amplitudes (see Fig. 2d). For a more quantitative analysis we plot the peak intensities $I(G_+)$ and $I(G_-)$ as a function of laser polarization (see Fig. 2c). Based on the relative shift of the $G_{+/-}$ peaks, we obtain a strain anisotropy compatible with a purely uniaxial stress and a Poisson ratio ν of about 0.15. Using $\Delta\varphi$ as a free parameter, we fitted $I(G_+)$ and $I(G_-)$ with (1) and obtained $\Delta\varphi \sim -1°$. This result is consistent with the hexagonal shape of the graphene flake visible in Fig. 2a, whose sides are expected to be oriented in the zigzag direction[24–26]. Here, it should be noted that strain engineering often requires a specific relative orientation between strain and the crystal axes[3] that can be directly tested by actuating the MAMs in the proper orientation.

The local traction of the MAMs has an important impact on the topography of the graphene/PMMA membrane, as illustrated in Fig. 3. Wrinkles and out-of-plane deformations are not unexpected[27] and are crucial for the prediction and design of custom strain profiles. Figure 3a shows the AFM topography of the sample after MAM actuation, superimposed to the strain profile calculated from the 2D Raman peak shifts. Wrinkles are observed in the region with large uniaxial strain, and occur at two qualitatively different length scales (see magnified scan in Fig. 3b and Supplementary Information): oscillations with periodicity ~ 750 nm and amplitude ~ 20 nm; oscillations with periodicity ~ 75 nm and amplitude of few nanometers. The former extend vertically up to the MAMs edge and involve the PMMA-covered regions, while the latter are confined to the polymer-free window. They can be better observed in Fig. 3c. As detailed in the Supplementary Information, compatibly with what we found comparing Raman spectroscopy data and finite-element method (FEM) simulations, these wrinkles emerge as a relaxation mechanism for a compressive stress component. In fact, numerical results indicate that MAM pulling leads to a lateral compression of the suspended membrane beyond the natural shrinkage due to the Poisson ratio, leading to compressive stress. We attribute larger wrinkles to the graphene/PMMA mechanical response, while smaller ones are directly linked to graphene. Recognizing the existence of these effects is crucial to properly interpret the FEM results. Importantly, this means that the formation and location of nano-wrinkles can be predicted based on the occurrence of compressive stress in the simulations.

A set of parameters determines the mechanical action of the MAMs. These include: adhesion between the various components of the device; viscoelastic properties of PMMA; contraction as a function of the e-beam dose. The observation of strain profiles on the SiN holes is indicative of a good PMMA-graphene adhesion in the exposed MAMs region, while graphene was found to be able to slide in the unexposed PMMA-covered areas. Experimental data are correctly reproduced by assuming an isotropic shrinkage of PMMA of ~ 0.6%, with a remarkable corresponding tensile stress of ~ 100 MPa and pulling force of about 10 μN per MAM, in the present geometry. This latter feature is in good agreement with the estimate made by Sameer et al.[28] Further details are available in the Supplementary Information. Viscoelastic reaction is directly linked to the crucial possibility of multiple contractions of the MAMs, which is investigated in Fig. 4. Similarly to real muscles, MAMs were indeed found to partially and slowly relax over time, depending on the amount of cross-linking. Most importantly, the graphene layer could be stretched again by a subsequent e-beam exposure: this is demonstrated in Fig. 4c, reporting the strain cross-section in a double pulling/relaxation cycle for a second sample. As-fabricated devices (step 1) were first exposed to e-beam radiation (step 2), then relaxed by heating the sample in at 100 °C for 90 min (step 3) and finally strained again by a second e-beam exposure (step 4). An example of relatively fast partial decay of the induced strain versus time is visible in Fig. 4b, where we report the time evolution of the strain caused by the MAMs. A characteristic exponential relaxation with a time constant ~ 12 min is obtained, with a residual strain ~ 0.3%. Different relaxation times were observed, spanning from tens of minutes up to more than a week, depending on the specific cross-linking procedure, with some degree of sample variability. We believe this is mainly due to cross-link heterogeneity, which can be relevant in highly cured PMMA[29] that leads to a broadened glass-transitions distribution as well as sensitivity to ambient fluctuations of the environmental conditions, such as humidity and temperature. This hypothesis is supported by the finding that the strain can be totally released if the sample is exposed to heat, a characteristic feature of glass-transition behaviour of cross-linked PMMA.

In conclusion, we have demonstrated that cross-linked PMMA domains can be used as effective actuators for the creation of controlled in-plane strain profiles and nano-wrinkle arrangements in graphene. We showed that the MAM technology, combined with the analysis techniques used in this work (μ-Raman, SEM, AFM, FEM) constitute a complete set of tools to manipulate and investigate the mechanics of layered materials on the nano-scale. By inducing a local uniaxial strain profile we determine the crystallographic orientation of the flake and demonstrate the creation of local nano-wrinkles. We showed that MAMs can be relaxed and contracted again, making it possible to perform multiple strain experiments on the same flake. As first examples, we believe that the present method could be used to investigate gauge fields in two-dimensional materials or investigate wrinkles and their impact on graphene chemical properties (e.g., on hydrogen adsorption)[30].

The data reported in the paper were obtained using three different membranes. The discussed method was further tested on about 30 membranes.

## Acknowledgements

We thank Vincenzo Piazza for the support in the implementation of the polarized Raman scattering set-up and Marco Polini, Giuseppe Grosso and Guido Menichetti for useful discussions. The research leading to these results has received funding from the European Union's Horizon 2020 research and innovation program GrapheneCore1 (GA 696656), the ERC Advanced Grant SOULMan (GA 321122) and from the GRAPHENE Flagship Project (contract NECT-ICT-604391). FC acknowledges financial support from Fondazione Silvio Tronchetti Provera.

## Author contributions.

The experiment was conceived and guided by FC, AP, PP and SR. FC, AP fabricated the devices. FC, AP and SR have performed the *in-situ* SEM studies. FC performed the Raman and AFM experiments. Simulations were done by FC and AP. CVD graphene flakes were grown by VM and CC. All the authors contributed to writing and editing the manuscript.

## Materials and Methods

**Graphene growth and transfer.** Single-layer, single-crystals graphene was grown on copper by Chemical Vapour Deposition (CVD) in a deterministic way using arrays of metallic nano-particles as nucleation points for the growth as described in ref.[31]. The spacing in between the crystals was 200 μm and each crystal had a size of ~ 150 μm (see Fig. S1). A 110±10 nm thick layer of AR-P 679.2 (PMMA) was spun on top of the copper foil used for graphene growth. This polymeric layer was used for the transfer as well as for the micrometric artificial muscles (MAM) actuation. Graphene was detached from copper using electrochemical delamination[26,32]. We first attached a PDMS frame to the copper/graphene/PMMA stack in order to handle the polymeric foil once released from the copper. The 2 mm-thick PDMS frame had a hole of 1 cm of diameter in the centre, which limits the graphene/PMMA region that will be transferred during deposition. After delamination, the graphene/PMMA was rinsed in DI water and deposited on the pre-patterned substrate with micrometric accuracy. In order to do that, a custom-made micromanipulator/microscope set-up was employed. After the substrate-graphene/PMMA contact, the sample was heated at 120 °C for 5 min to increase the graphene-substrate adhesion.

**Substrate fabrication.** The substrate consists of a 250 μm commercial silicon wafer (from Si-Mat) sandwiched between two 300 nm thick stoichiometric silicon nitride layers deposited with PECVD at high temperature (~ 850 °C) in order to have a strong residual tensile stress at room temperature. On one side (the "back") of the chips diced from the wafer, 0.6×0.6 mm$^2$ square windows were opened on the $Si_3N_4$ layer by using S1818 resist as a mask, UV-lithography for pattern definition and finally $CF_4$-based plasma etching. Following this step, an anisotropic wet etching of Si (KOH 30% at 70 °C) was performed, leaving only 2 to 5 μm of Si underneath the $Si_3N_4$ of the other side (the "front") of the chips for improved membrane support before the final release. The front side was then patterned in correspondence of the membranes defined in the back side using AR-P 6200 (CSAR) resist as a mask, 30 keV e-beam lithography and finally dry etching. The geometry chosen was made of arrays of 10 μm diameter circular holes as well as markers for the next e-beam irradiation steps. The remaining Si was then etched, suspending the patterned 300 nm $Si_3N_4$ membranes. Samples were then coated in Ti/Au (10/50 nm) to ground the sample and minimize charging effects. After graphene deposition, the devices were ready for MAMs actuation.

**MAMs actuation and characterization.** A region of PMMA was removed in the graphene/PMMA stack for undisturbed access to the graphene layer. This could be done following two alternative approaches. A first option was to use a conventional aligned e-beam lithographic step (300 μC/cm$^2$ at 10 keV) followed by resist development. A more general approach that can be extended to a wider set of polymers is to mildly crosslink the film using a more "robust" exposure (10 mC/cm$^2$ at 10 keV) and dissolve the unexposed polymer in its solvent (in our case, acetone). Both methods were found to be suitable for MAMs operation. A subsequent aligned e-beam exposure at much higher dose was performed to actuate the MAMs. The beam energy was reduced to exploit the increased cross section for electron scattering. Different energies were used in the range from 2 to 5 keV: all devices reported in the paper were obtained using 5 keV. We found the dose necessary to get a sizable amount of MAMs contraction slightly varying from sample to sample, in a range from 30 to 100 mC/cm$^2$s. After irradiation, samples were characterized using a micro-Raman system from Renishaw where we included the possibility to control the polarization state of the input and output laser beams. The 532 nm laser was focused through a 100x microscope objective with a corresponding laser spot with a ~ 500 nm diameter. We tested the impact of heating due to laser irradiation by changing integration time and laser power while monitoring the time evolution of the strain relaxation, using a non-uniform data sampling over time. By increasing the laser power, we observed an increase of the strain relaxation rate. At high laser power and sufficiently long integration time, a broadening of the 2D Raman peak is expected if thermal heating is not negligible, since the peaks are shifting during the sampling. By contrast, at the low laser power employed in our experiment (down to 50 μW) we never detected any broadening even if spanning over different integration times, indicating that the measurement does not impact strain relaxation dynamics. Wrinkles characterization was performed employing a commercial Icon Bruker AFM in *PeakForce Tapping mode* with standard *ScanAsyst-Air* tips. A typical nominal force of 1 nN was used to scan the samples.

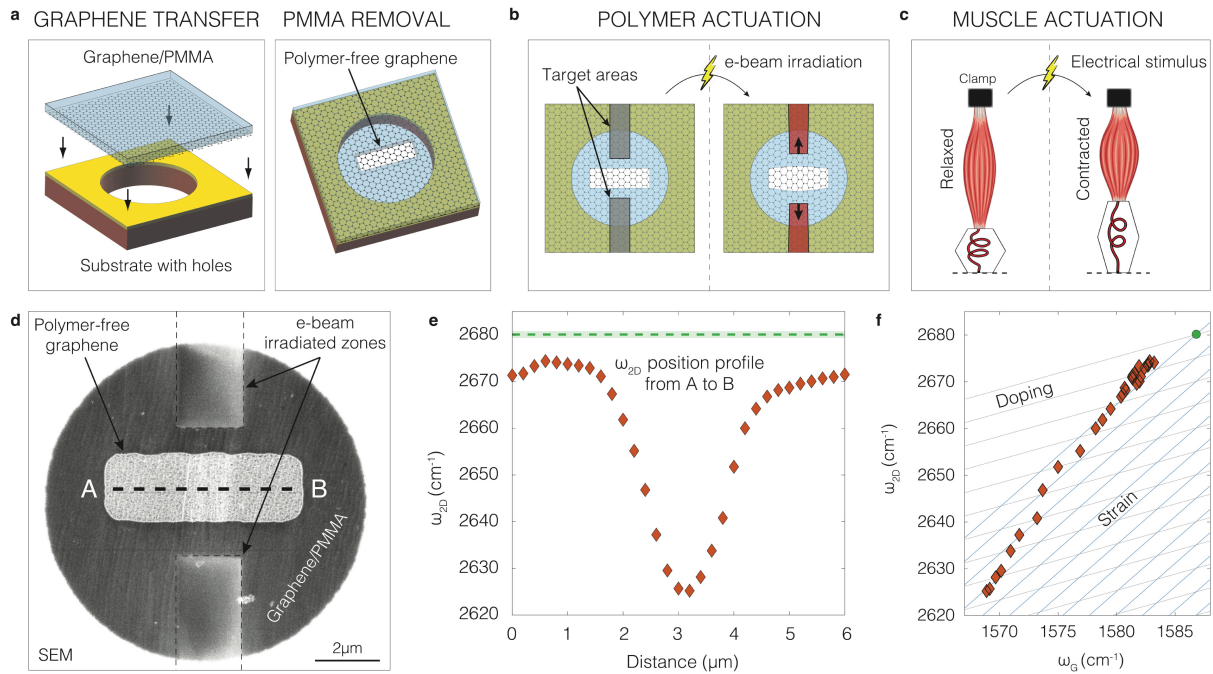

**Figure 1 – Graphene strain engineering platform. a.** Devices are fabricated by depositing CVD graphene on SiN membranes patterned with circular holes with a diameter of 10 µm. A 100 nm-thick 950K PMMA layer is used as the transfer vector and patterned to release part of the graphene membrane. **b.** Graphene can be pulled by e-beam irradiation of suitably chosen graphene/PMMA adjacent regions. This induces a lateral shrinkage of PMMA and, in turn, strain in graphene central portion. **c.** The action of the PMMA can be assimilated to the one of an artificial muscle, which contracts in response to an electrical stimulus. **d.** Scanning electron micrograph of one of the studied devices. Polymer-free graphene is visible as a horizontal bright rectangle at the centre of the circular SiN hole. The e-beam irradiated graphene/PMMA regions are visible as grey rectangles at the top and bottom. **e.** MAMs excitation leads to a marked shift in the 2D Raman peak. Data refer to the cross-section AB in panel (**d**) after the excitation of the MAMs. At the maximum pulling point, the 2D peak is shifted to 2625 cm$^{-1}$, corresponding to a strain of 0.8%. **f.** The origin of the shift is clearly related to strain and not to doping, as visible from the correlated evolution of the 2D and G peaks: data align consistently with the strain-driven slope (cyan). The green dashed line in panel (**e**) and the green spot in panel (**f**) correspond to the Raman peak positions at zero-strain.

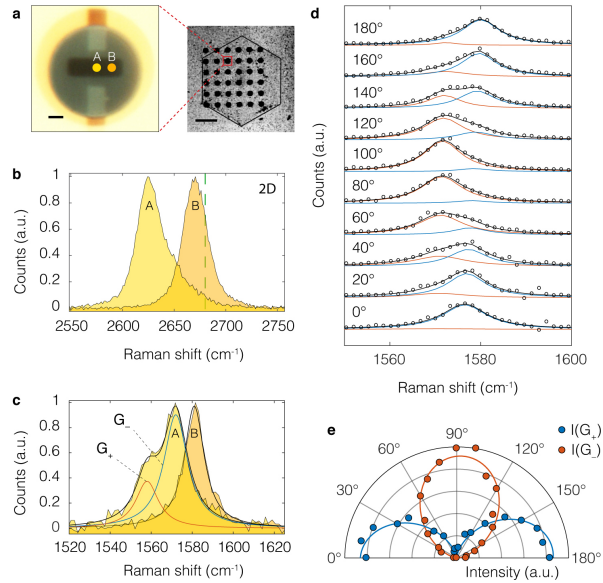

**Figure 2 – Evidence for uniaxial strain. a.** Optical micrographs of one of the studied devices (left) with the relative zoom-out micrograph of the sample (right) where are visible the 6x6 array of holes in the SiN and the deposited graphene single-crystal (highlighted by the hexagon). Scale bars correspond to 2 μm (left) and 50 μm (right) **b.** The 2D peak measured at the centre of the membrane (yellow dot in panel (**a**)) is significantly shifted with respect to the one obtained 2 μm away (orange dot), in a region where no significant strain is expected. **c.** The evolution of the G peak is more complex, with the formation of a clear shoulder which is consistent with the splitting of the $G_+$ and $G_-$ Raman modes, as expected in the presence of anisotropic strain. **d.** Polarized micro-Raman is used to separate the contribution of the $G_+$ and $G_-$ peaks, whose amplitude is modulated by the relative angle between the input and output light polarization. **e.** Polar plot of the two modes amplitudes. A relative rotation of about $\Delta\varphi \sim -1°$ between the zigzag axis and the strain axis is obtained by fitting the two oscillations. This orientation is consistent with the flake boundaries highlighted in panel (**a**).

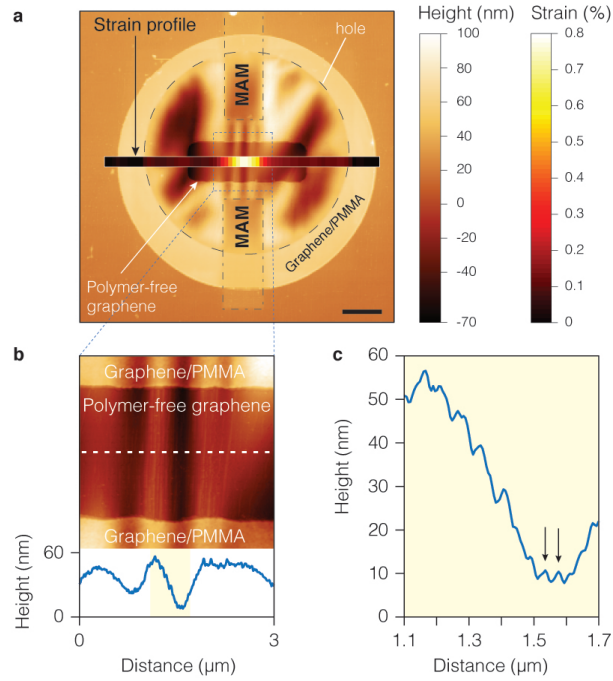

**Figure 3 – Topography and strain-induced wrinkles. a.** Topographic map of one of the studied samples. The dashed circle indicates the edge of the SiN hole. The PMMA corresponds to the bright yellow circular region that covers the hole and overlaps with the SiN substrate on a 2 μm-wide ring. The central rectangular slit corresponds to the polymer-free graphene that is pulled by the MAMs. The latters can be seen as flat vertical rectangles on the top and bottom parts of the scan field. The suspended central region displays a sizeable out-of-plane distortion (on the order of tens of nanometers). The creation of wrinkles is linked to the relaxation of compressive strains caused by the local anisotropic pulling of the MAMs (see Supplementary Information). The strain profile is shown in overlay in order to highlight the position of the stretched graphene region. Scale bar: 2 μm. **b.** Various wrinkles are visible in correspondence of strained graphene: large wrinkles with a periodicity ∼ 750 nm are plausibly associated with the mechanical response of PMMA; smaller wrinkles with a periodicity of ∼ 75 nm are only visible in the polymer-free graphene and are linked to the mechanical response of graphene. **c.** Magnified cross-section of panel (**b**), highlighting the two different oscillations that are observed on graphene.

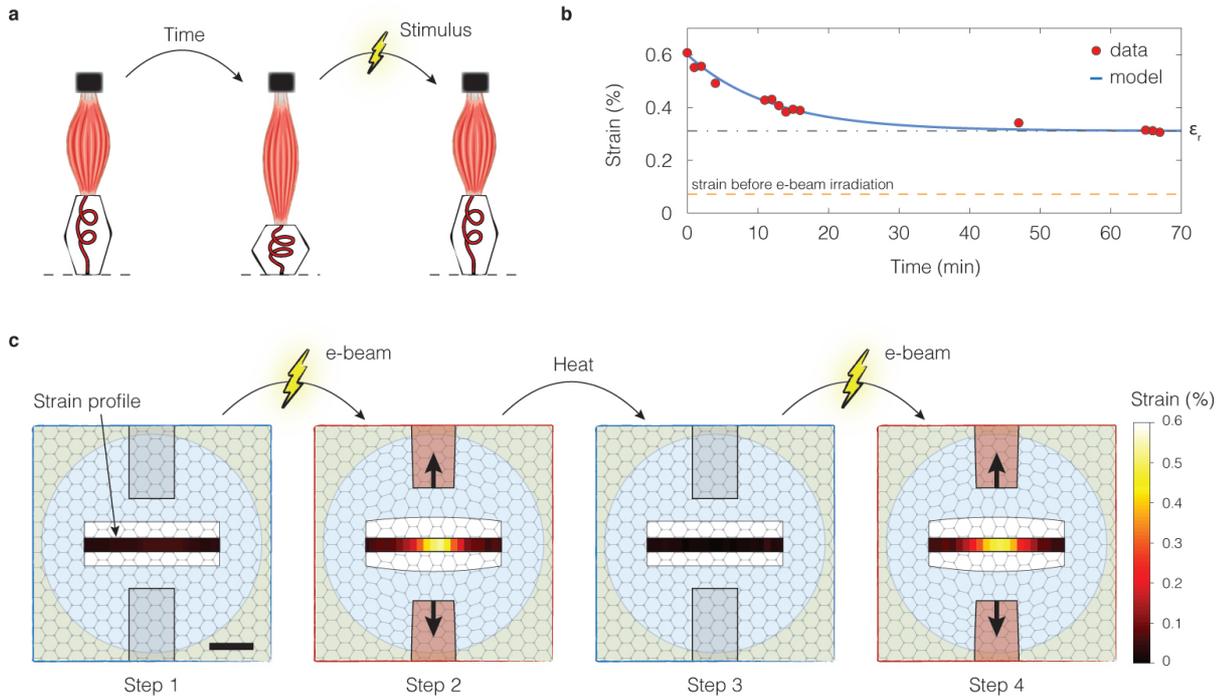

**Figure 4 – Relaxation and re-contraction of the MAMs. a.** MAMs contract when excited by electron irradiation and can relax as a function of time or due to exposure to heat. A second stimulation is able to re-contract the actuator. **b.** The time evolution of MAM traction can be deduced by the drift over time of the graphene strain in the region pulled by the MAMs. Experimental data (red spots) are well described by a simple viscoelastic relaxation law leading to an exponential decay (blue line) with a characteristic time of 12 minutes. A residual strain ($\varepsilon_r$) of 0.3% is obtained asymptotically. **c.** Full relaxation is obtained when the MAMs are exposed to heat. The as-fabricated sample (step 1) is first pulled by actuating the MAMs (step 2), then relaxed by heating to 100 °C for 90 min (step 3), and finally pulled again (step 4). As visible from the experimental data in overlay, strain can be fully released by relaxation step 3. The second e-beam exposure in step 4 re-contracts the MAMs and the resulting strain profile is comparable to the one obtained on the first contraction in step 2. The strain profiles share the same colour bar and scale bar (2 µm).

9